\documentclass[preprint,superscriptaddress,
amsmath,amssymb,aps,showkeys,showpacs,
final,secnumarabic,nofootinbib]{revtex4-2}

\usepackage{cmap}
\usepackage[T1,T2A]{fontenc}
\usepackage[utf8]{inputenc}
\usepackage[russian,english]{babel}
\usepackage{color}
\usepackage{graphicx}% Include figure files
\usepackage{dcolumn}% Align table columns on decimal point
\usepackage{bm} % bold math
\usepackage[unicode=true,colorlinks=true,linkcolor=magenta, urlcolor=blue, citecolor = blue,breaklinks]{hyperref}
\usepackage{multirow}
\usepackage{url}
\usepackage{breakurl}
\DeclareGraphicsExtensions{.eps}

\begin{document}

\title{Analysis of the TAIGA-HiSCORE Data Using the Latent Space of Autoencoders} 

\def\addressa{Lomonosov Moscow State University, Skobeltsyn Institute of Nuclear Physics. Moscow, 119991 Leninskie gory, 1, bld. 2, Russia}
\def\addressb{Research Institute of Applied Physics. Irkutsk, 664003, blv. Gagarina, 20, Russia}
\def\addressc{Institute for Informatics and Automation Problems of the National Academy of Science of the Republic of Armenia. Yerevan, 0014, P. Sevak str., 1, Republic of Armenia}

\author{\firstname{Yu.Yu.}~\surname{Dubenskaya}}
\email[E-mail: ]{dubenskaya@theory.sinp.msu.ru}
\affiliation{\addressa}
\author{\firstname{S.P.}~\surname{Polyakov}}
\affiliation{\addressc}
\author{\firstname{A.P.}~\surname{Kryukov}}
\affiliation{\addressa}
\author{\firstname{A.P.}~\surname{Demichev}}
\affiliation{\addressa}
\author{\firstname{E.O.}~\surname{Gres}}
\affiliation{\addressb}
\author{\firstname{E.B.}~\surname{Postnikov}}
\affiliation{\addressa}
\author{\firstname{A.Yu.}~\surname{Razumov}}
\affiliation{\addressa}
\author{\firstname{P.A.}~\surname{Volchugov}}
\affiliation{\addressa}
\author{\firstname{D.P.}~\surname{Zhurov}}
\affiliation{\addressb}

\begin{abstract}
The aim of extensive air shower (EAS) analysis is to reconstruct the physical parameters of the primary particle that initiated the shower. The TAIGA experiment is a hybrid detector system that combines several imaging atmospheric Cherenkov telescopes (IACTs) and an array of non-imaging Cherenkov detectors (TAIGA-HiSCORE) for EAS detection. Because the signals recorded by different detector types differ in physical nature, the direct merging of data is unfeasible, which complicates multimodal analysis. Currently, to analyze data from the IACTs and TAIGA-HiSCORE, a set of auxiliary parameters specific to each detector type is calculated from the recorded signals. These parameters are chosen empirically, so there is no certainty that they retain all important information contained in the experimental data and are the best suited for the respective problems. We propose to use autoencoders (AE) for the analysis of TAIGA experimental data and replace the conventionally used auxiliary parameters with the parameters of the AE latent space. The advantage of the AE latent space parameters is that they are not biased by pre-established assumptions and constraints and still contain in a compressed form the physical information obtained directly from the experimental data. This approach also holds potential for enabling seamless integration of heterogeneous IACT and HiSCORE data through a joint latent space. To reconstruct the parameters of the primary particle of the EAS from the latent space of the AE, a separate artificial neural network is used. In this paper, the proposed approach is used to reconstruct the energy of the EAS primary particles based on Monte Carlo simulation data for TAIGA-HiSCORE. The dependence of the energy determination accuracy on the dimensionality of the latent space is analyzed, and these results are also compared with the results obtained by the conventional technique. For signals recorded by TAIGA-HiSCORE, it is shown that when using the AE latent space, the energy of the primary particle is reconstructed with satisfactory accuracy.
\end{abstract}

\keywords{machine learning, autoencoder, latent space, gamma astronomy, energy reconstruction \\[5pt]}

\maketitle

\section{Introduction}\label{intro}

Machine Learning (ML) is currently one of the most promising areas of computer science, experiencing rapid growth and significant impact across various industries.
AEs~\cite{AE-about} are a powerful and versatile tool in the field of ML, with a wide range of applications, particularly for generative modeling, anomaly detection, image denoising, dimensionality reduction, and feature extraction. An AE is a type of neural network architecture designed to efficiently compress input data by encoding it into a lower-dimensional latent space and then decoding it back to the original dimensions. By definition, the latent space of the AE is the input to the decoder, but this is not its only possible application. Using the AE latent space as input for another neural network is an effective technique in various ML applications. This approach leverages the ability of AEs to learn a compressed, meaningful representation of the input data, which can then serve as a more robust and lower-dimensional input feature set for subsequent tasks. Solutions based on this approach are applied in a wide range of areas of science and engineering, such as determining the severity of Parkinson's disease~\cite{AE-for-med}, improving autonomous navigation of drones~\cite{AE-for-drones}, and classifying partially labeled datasets~\cite{AE-for-classification}.

In this paper, we propose to use the latent space of the AE as a set of essential features to determine the energy of the primary particle that produced the EAS when analyzing the data of the TAIGA-HiSCORE detector array~\cite{HiSCORE-about} from the TAIGA experiment. TAIGA-HiSCORE is a large-scale array of wide-angle Cherenkov detectors (stations), designed to detect and study EASs from primary high-energy cosmic rays and gamma rays. The non-ML energy reconstruction technique currently in use is based on extracting auxiliary parameters from the recorded data and using these parameters to solve the problem. We assume that such an approach may lead to the loss of useful information from the experimental data, since the set of auxiliary parameters is fixed in advance, as well as the algorithm for calculating them based on the recorded signals. Using AE allows us to build a similar set of parameters, but to optimize the values of these parameters based on the available training data. We can also vary the dimensionality of the latent space, thereby achieving an optimal representation of the physics of the phenomenon.

We believe that the proposed approach is promising specifically for TAIGA-HiSCORE data for several reasons. First, the data are heterogeneous: each TAIGA-HiSCORE station registers Cherenkov photons and the signal arrival time. Also there could be stations that failed to capture the signal. This information must be explicitly taken into account in the analysis and distinguished from the case where the station was not triggered because there was no signal. ML techniques, including AEs, cope well with such heterogeneous data by introducing multiple input channels, but this means increasing the input dimensionality. The latent space of the trained AE used as input to solve specific physics problems such as energy reconstruction will reduce the computational cost of these problems. Secondly, the data are noisy. We expect that using the latent space of the AE will reduce the noise and improve the accuracy of the results when reconstructing the energy and other parameters of the shower. Thirdly, the latent space of the trained AE opens up new opportunities in multimodal analysis by providing a unified framework for combining data from different installations. Despite the significant difference in apertures, the imaging telescopes and TAIGA-HiSCORE are able to register signals from the same EASs~\cite{TAIGA-joint-events}. Analysis of such joint events is a way to further improve the accuracy of energy reconstruction. However, the data obtained from the telescopes differ in physical meaning, signal values, and dimensionality from the TAIGA-HiSCORE data. AEs offer a powerful solution for reducing and preprocessing data from IACT and TAIGA-HiSCORE to ensure seamless integration of these heterogeneous data sources.
This approach is further motivated by the proven efficacy of ML-based methods for processing astrophysical data~\cite{ML-IACT-overview}, in particular IACT~\cite{Sim-TAIGA-IACT,CGAN-for-Gen} and TAIGA-HiSCORE data~\cite{CNN-for-HiSCORE,FCNN-for-HiSCORE}.  

\section{\label{sec:taiga-hiscore}TAIGA-HiSCORE}

\begin{figure}
\includegraphics[width=8cm]{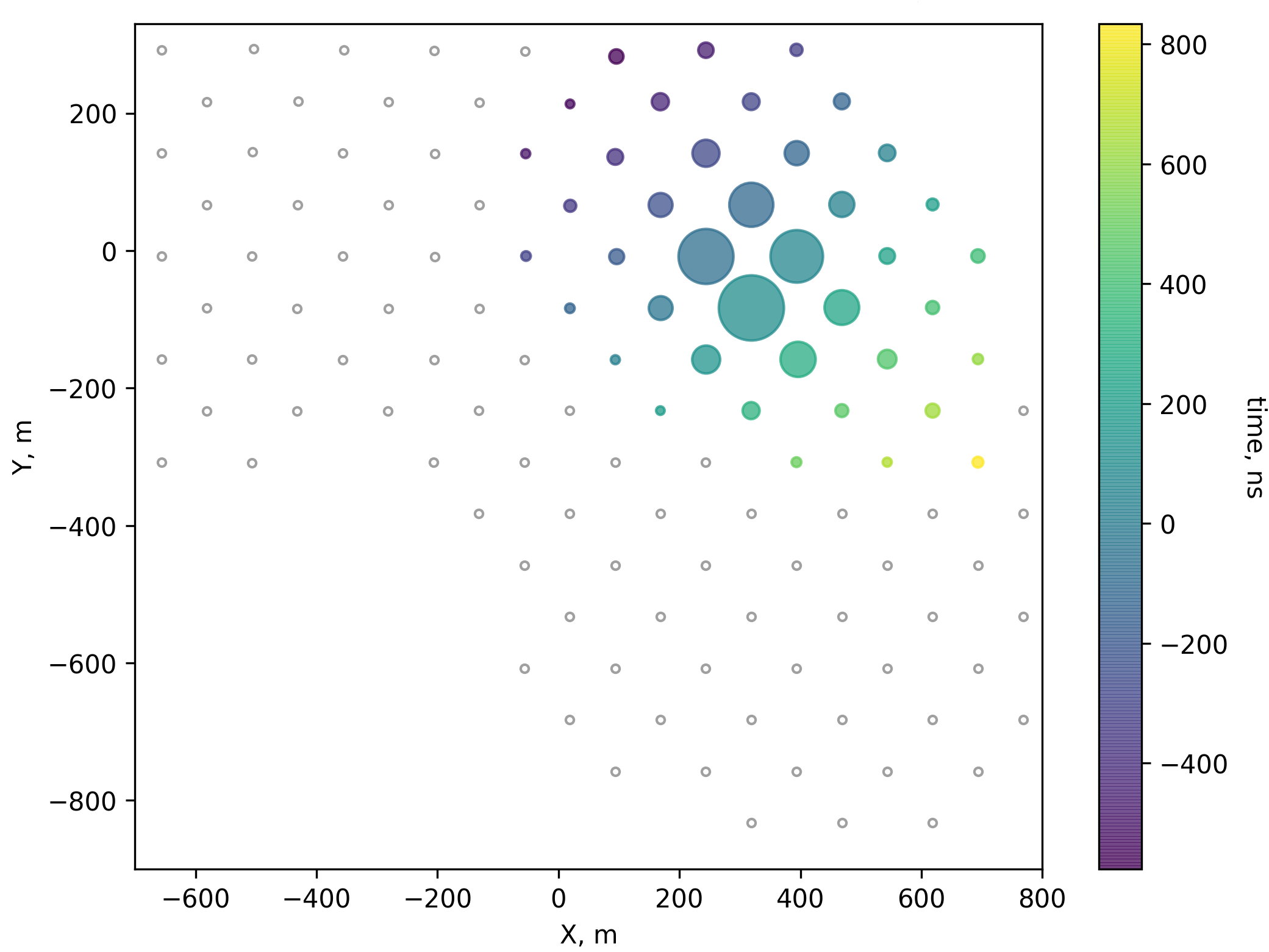}
\caption{\label{fig:hiscore-event-image} A visualization of a TAIGA-HiSCORE event. Each point corresponds to a HiSCORE station, the color from purple to yellow indicates the time of signal recording, the size indicates the signal amplitude, and small gray dots indicate stations that were not triggered.}
\end{figure}

The TAIGA-HiSCORE stations measure and record the amplitude and the arrival time of the Cherenkov radiation front from EAS~\cite{TAIGA-Spectrum}. 
The TAIGA-HiSCORE array currently includes 121 stations whose positions (see Fig.~\ref{fig:hiscore-event-image}) can be approximated by a rectangular grid. The data for a single event can therefore be represented by a set of two-dimensional arrays whose values encode heterogeneous data including arrival times and amplitudes from the corresponding HiSCORE stations. For the purposes of convolutional neural networks, such set is equivalent to a multi-channel image, so we call them quasi-images.

The currently used technique for reconstructing and determining the energy of the EAS is described in detail in~\cite{Current-method}. In brief, it relies on the following auxiliary physical parameters that are derived from the recorded data:
\begin{itemize}
  \item pulse shape analysis: rise time and width of the arrival time distribution;
  \item geometry of shower development: arrival time front curvature;
  \item lateral distribution function: signal amplitude vs. core distance;
  \item Cherenkov light density: flux at 200m from shower core.
\end{itemize}
Based on these auxiliary parameters, empirical formulas were selected, using which the EAS parameters (in particular, the energy of the primary particle) can be reconstructed with acceptable accuracy. Despite its relative effectiveness, this reconstruction technique remains simulation-dependent, having been empirically optimized and validated through Monte Carlo simulations. A feature of this technique is that each auxiliary parameter has a physical meaning. However, these parameters are chosen intuitively, their number is determined and cannot be arbitrarily increased.

\section{\label{sec:ml-approach}ML-based approach to energy reconstruction}
\subsection{General architecture of the system}
\begin{figure}
\includegraphics[width=16cm]{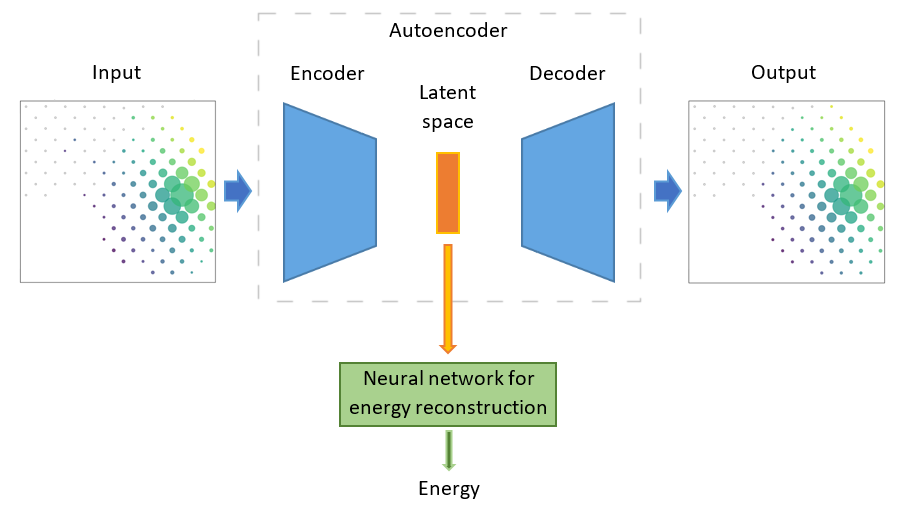}
\caption{\label{fig:common-scheme} The general architecture of the proposed model. First, the AE is trained to reconstruct the registered TAIGA-HiSCORE data, then the energy reconstruction network is trained to reconstruct the energy from the AE latent space.}
\end{figure}

We propose an alternative approach to determining the EAS parameters from the TAIGA-HiSCORE data: to replace the currently used auxiliary physical parameters calculated using pre-determined algorithms with parameters extracted from the data automatically by the encoder of an AE trained on similar data.
The quality of the extracted parameters can be optimized by varying hyperparameters, such as AE architecture, the dimensionality of the latent space, loss function etc. The latent parameters of the AE can then be considered as essential features of the input data that eliminate noise but contain as much meaningful information as possible in a compressed form. It should be noted that after training the AE, to obtain the latent space we only need the encoder. We also propose to train an auxiliary neural network that will reconstruct the energy of the primary particle, receiving the latent space (the essential features) as input. This approach opens a path to a fully data-driven reconstruction pipeline, with future extensions to particle type, direction and any other EAS parameters. The architecture of the proposed model is shown in Fig.~\ref{fig:common-scheme}.

Since we train the AE and the energy reconstruction network independently, we will discuss them separately in more detail below.

\subsection{Autoencoder}
\subsubsection{Preparation of the training data}
Initially, we took 97415 TAIGA-HiSCORE events simulated with the Monte Carlo method. 
The primary particles were gamma quanta, with parameters taking values in the following ranges: $100\ \mathrm{TeV} \leq E \leq 1000\ \mathrm{TeV}$, $30^{\circ} \leq \theta \leq 40^{\circ}$, $120^{\circ} \leq \phi \leq 240^{\circ}$, where $E$ is the particle energy and $\theta$ and $\phi$ are the zenith and azimuth angles of the shower axis, respectively.
The input data for each of the 121 stations included the amplitude (the number of photoelectrons) $A$, the average $t$ and the approximated standard deviation $s$ of photoelectron detection times.

Several restrictions excluding less informative stations and events were applied.
For stations, the trigger threshold was set to 100 photoelectrons. 
Additionally, the maximum range of average detection times for each event was set to 3 microseconds, with stations discarded until the requirement was satisfied.
After that, events with fewer than 10 triggered stations were excluded from the set, resulting in 79745 events with 42 triggered stations on average.
Finally, we removed events which had the highest amplitude station located at the edge of the station array, resulting in 29314 events with 56 triggered stations on average.

Station positions used in the Monte Carlo simulation were approximated by a rectangular $17 \times 12$ grid with 106 m between adjacent stations. 
The average distance between a station position and the corresponding grid node was 0.78 m; the maximum distance was 10 m.

To prepare the AE input data for each event, the data from the triggered stations were arranged in three $17 \times 12$ arrays for $A$, $t$, and $s$, and an additional array with indicators showing whether the given station has been triggered (trigger indicators).
Thus, input data for each event were a $17 \times 12 \times 4$ quasi-image.
For the grid positions which did not have a corresponding station or whose corresponding station was not triggered, all input values were set to 0. 

The input data were augmented using reflection along the grid axes, producing 4 quasi-images for each Monte Carlo event.

\subsubsection{Autoencoder architecture}
Initially, variational autoencoders~\cite{VAE-about} (VAEs) were chosen to extract essential features from TAIGA-HiSCORE data.
Instead of single vectors of latent parameter values corresponding to each data instance, encoders in VAEs generate distributions that can overlap.
The overlap is controlled by a loss function parameter.
Preliminary experiments did not show significant advantages in using values of this parameter that forced significant overlap.
Therefore, at the moment, we use only the zero value of the parameter, at which latent parameter variations are negligible and our VAE effectively becomes a regular AE.
For inference events, we only calculate the average values $\mu$ from the encoder output and ignore the $\log \sigma^2$ outputs.

The VAE consists of a convolutional encoder and a transposed convolutional decoder without fully connected layers, mostly symmetrical to each other.
The encoder inputs are quasi-images and the decoder inputs are latent parameters.
As an additional input, each component receives a $17 \times 12 \times 3$ tensor of coordinate offsets (i.e. the differences along the $x$, $y$, and $z$ axes between the station positions and the corresponding grid node positions).
Each of the two components has approximately 750 thousand trainable parameters.

\begin{figure}
\includegraphics[width=13cm]{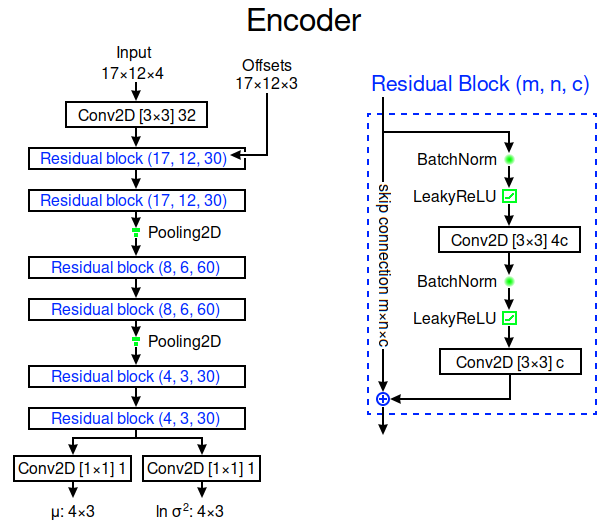}
\caption{\label{fig:encoder-architecture} The architecture of the encoder.}
\end{figure}

The encoder architecture for $12$ latent parameters is shown in Fig.~\ref{fig:encoder-architecture}.

The encoder output is two tensors interpreted as mean values and logarithms of variance of the latent parameters.
The decoder output is a $17 \times 12 \times 4$ tensor interpreted as a quasi-image except for the values corresponding to trigger indicators, to which the Heaviside step function $H$ with $H(0)=0$ must be applied first.

\subsubsection{Autoencoder training}
The AEs were trained using the loss function:
$$
L = C_A L_A + C_t L_t + C_s L_s + C_{\mathrm{trigger}} L_{\mathrm{trigger}}
$$
where $L_{\mathrm{trigger}}$ is binary cross-entropy, $L_A$ and $L_s$ are masked MSE, and $L_t$ is masked MSE multiplied by $\log_{10} A$.
The $C$ coefficients were optimized during our preliminary experiments and have the following values: $C_t = 10$, $C_A = 1$, $C_s = 0.5$, $C_{\mathrm{trigger}} = 0.0005$.

The trigger indicators of the input event are used as the mask.
This allows the AE to make guesses about the $A$, $t$, and $s$ values for the non-triggered stations without penalty for deviating from zero values used in the input.

The stations with higher signal amplitude are given more relative weight in $L_t$ because their detection times are less affected by noise and random fluctuations, making them more important for reconstruction of physical parameters.

The training set consisted of 58624 quasi-images prepared from 14670 Monte Carlo events.
Each AE was trained for 475 epochs with learning rate $10^{-3}$ followed by 25 epochs with lower learning rates.
During training, a second augmentation was applied: stations were masked randomly at each epoch with the probability inversely proportional to the signal amplitude.
For the first 450 epochs, approximately 37.5\% stations were masked, and for the remaining 50 epochs, the masking probability was decreased by a factor of 10. 

\subsubsection{Results of the TAIGA-HiSCORE data reconstruction}
The results for the trained AEs were compared on 2960 validation events. 
To select the optimal architecture, the values of the coefficients of determination ($R^2$) of the reconstructed quasi-images were compared depending on the dimensionality of the latent space $N$ of the AE. The results are shown in Table~\ref{tab:table1}. In addition to assessing the quality of image reconstruction, all these AEs were tested in solving the problem of primary particle energy reconstruction. A comparison of energy reconstruction results for AEs with different $N$ will be given in Section~\ref{sec:NN-energy}.

\begin{table*}[htbp]
\renewcommand{\arraystretch}{1.25}
\renewcommand{\tabcolsep}{5pt}
\begin{center}\caption{$R^2$ coefficients for restored quasi-images depending on the dimensionality of the latent space $N$}
\begin{tabular}{|c|c|c|c|c|c|c|}\hline
 & $N$=4 & $N$=6 & $N$=8 & $N$=12 & $N$=16  & $N$=20 \\ \hline
amplitude & 0.83241 & 0.91355 & 0.96229 & 0.97174 & 0.97374 & 0.9735  \\ \hline
time & 0.99922 & 0.99973 & 0.99973 & 0.99975 & 0.99976 & 0.99976  \\ \hline
standard deviation of times & 0.514 & 0.53314 & 0.54794 & 0.58936 & 0.60801 & 0.62623  \\ \hline
trigger & 0.71928 & 0.744 & 0.74836 & 0.75272 & 0.76456 & 0.78352  \\ \hline
\end{tabular}\label{tab:table1}
\end{center}
\end{table*}

Based on the comparison results, the option $N=12$ was selected, since smaller values of $N$ gave worse results of $R^2$, and larger values of $N$ did not lead to significant improvement.

For the AE with 12 latent features, the mean absolute errors are: 97.1 photoelectrons (amplitude), 5.87 ns (time), 3.91 ns (time standard deviation), and 10.1 stations (trigger).

\begin{figure}
\includegraphics[width=12cm]{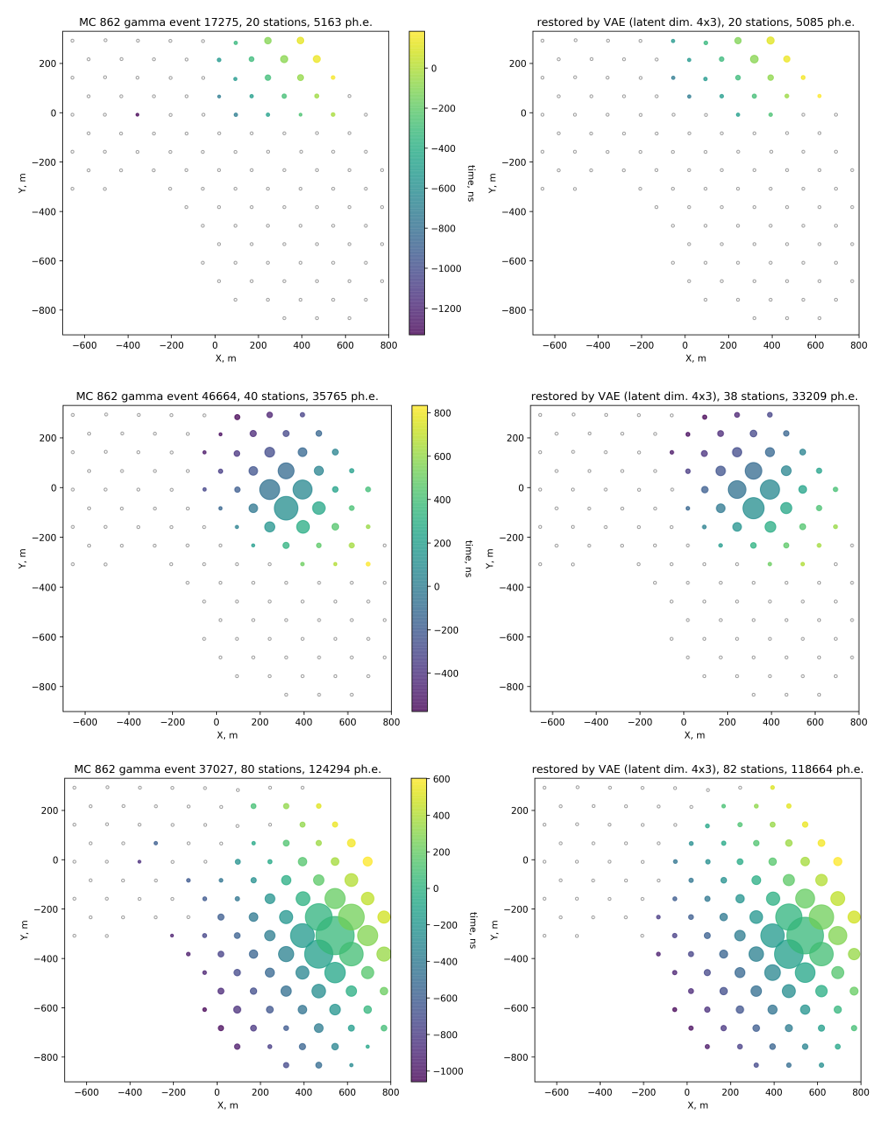}
\caption{\label{fig:encoder-results} Image reconstruction results for three events with different numbers of triggered stations. The original quasi-image is shown on the left, and its corresponding reconstructed image is on the right. Each image pair corresponds to a single event. The number of triggered stations is 20 for the first event, 40 for the second, and 80 for the third.}
\end{figure}

Fig.~\ref{fig:encoder-results} shows the results of the reconstruction of several input quasi-images with 20, 40, and 80 triggered stations by the AE with $N=12$.

\subsection{\label{sec:NN-energy}Neural network for energy reconstruction}

\begin{figure}
\includegraphics[width=10cm]{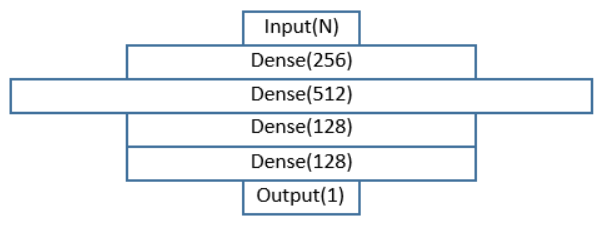}
\caption{\label{fig:NN-architecture} The architecture of the network for energy reconstruction.}
\end{figure}

The neural network for energy reconstruction is a multilayer perceptron. The input size of the network corresponds to the latent space dimensionality $N$ of the AE, for which we tested values of 4, 6, 8, 12, 16, and 20. The loss function is the mean square error (MSE). The output data is the energy of the primary particle (1 value). The architecture of the network is shown in Fig.~\ref{fig:NN-architecture}.

\begin{figure}
\includegraphics[width=12cm]{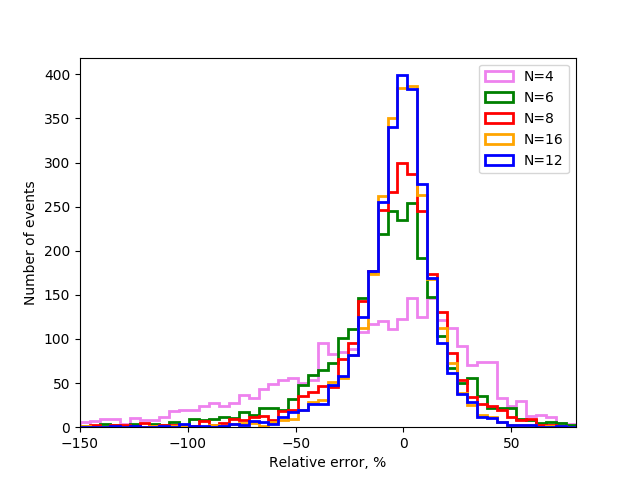}
\caption{\label{fig:NN-results-total} The distribution of the relative energy reconstruction error for different dimensionalities of the AE latent space. The relative energy reconstruction error is calculated as the ratio of the difference between the real and reconstructed energies to the real energy value, in percent.}
\end{figure}

The network was trained on 35,500 sets of latent features along with the energy values of the corresponding Monte Carlo gamma quanta events.
Although the choice of the dimensionality of the latent space $N$ was mainly based on the image reconstruction results, the energy reconstruction results were also taken into account. Fig.~\ref{fig:NN-results-total} shows the distribution of the relative error of energy reconstruction for different values of $N$ ($N = 4, 6, 8, 12,$ and $16$). The relative energy reconstruction error is calculated as the ratio of the difference between the real and reconstructed energies to the real energy value, in percent. The results for $N=20$ are not substantially different from those for $N=12$ and $N=16$. The means and standard deviations for the relative energy reconstruction error depending on $N$ value are presented in Table~\ref{tab:table2}.

\begin{table*}[htbp]
\renewcommand{\arraystretch}{1.25}
\renewcommand{\tabcolsep}{15pt}
\begin{center}\caption{The means and standard deviations for the relative energy reconstruction error for different $N$ values}
\begin{tabular}{|c|c|c|c|c|c|}\hline
 & $N=4$ & $N=6$ & $N=8$ & $N=12$ & $N=16$  \\ \hline
mean(Relative error), \%  & -20.4  & -9.4 & -4.7 & -2.8 & -2.4  \\ \hline
std(Relative error), \%  & 52.5  & 34.4 & 25.1 & 18.9 & 19.1  \\ \hline
\end{tabular}\label{tab:table2}
\end{center}
\end{table*}

\begin{figure}
\includegraphics[width=12cm]{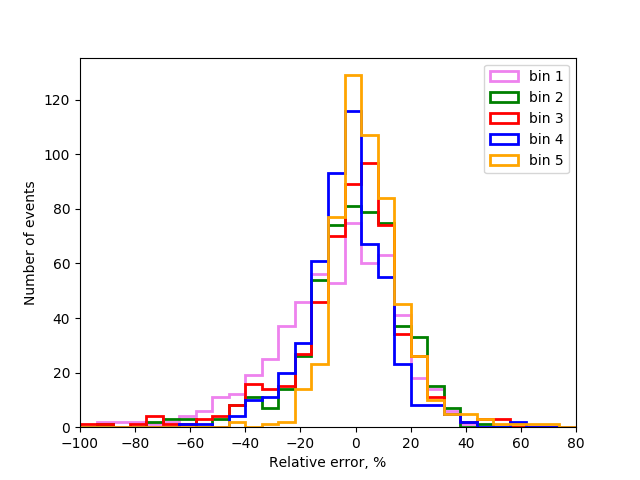}
\caption{\label{fig:NN-results-p12} The distribution of the relative energy reconstruction error for $N=12$ for 5 energy bins (bin 1: 100--160 TeV, bin 2: 160--260 TeV, bin 3: 260--420 TeV, bin 4: 420--650 TeV, bin 5: 650--1000 TeV).}
\end{figure}

Thus, the energy reconstruction results are consistent with the image reconstruction results. The optimal dimensionality of the latent space was found to be 12, since further increase in $N$ does not lead to a significant decrease in the energy reconstruction error.
After fixing the latent dimensionality at 12, we examined relative energy reconstruction errors across several energy ranges, called energy bins. The bin widths are chosen so that each bin contains approximately the same number of events. The distributions of the relative energy reconstruction error for different bins
is shown in Fig.~\ref{fig:NN-results-p12}, while the means and standard deviations are presented in Table~\ref{tab:table3}.

\begin{table*}[htbp]
\renewcommand{\arraystretch}{1.25}
\renewcommand{\tabcolsep}{3pt}
\begin{center}\caption{The means and standard deviations for the relative energy reconstruction error for $N=12$ for different energy bins}
\begin{tabular}{|c|c|c|c|c|c|}\hline
\rule{0pt}{25pt}  & \shortstack{bin 1: \\ 100--160 TeV}
  & \shortstack{bin 2: \\ 160--260 TeV}
  & \shortstack{bin 3: \\ 260--420 TeV}
  & \shortstack{bin 4: \\ 420--650 TeV}
  & \shortstack{bin 5: \\ 650--1000 TeV}  \\ \hline
mean(Relative error), \%  & -8.1  & -2.2 & -2.3 & -2.7 & 4.7  \\ \hline
std(Relative error), \%  & 24.2  & 19.2 & 20.4 & 14.7 & 13.1  \\ \hline
\end{tabular}\label{tab:table3}
\end{center}
\end{table*}

As can be seen, the model systematically overestimates low energies and underestimates high ones. Nevertheless, these results are consistent with those of the conventional non-ML-based reconstruction technique. Thus, in~\cite{Conventional-energy-results2016}, the accuracy of the energy reconstruction is given as 10--15\% without specifying the exact energy range for which these values are determined. In~\cite{Conventional-energy-results2013}, the relative error was estimated using Monte Carlo simulation and was about 20\% for the 100 TeV and improved to 10\% for 1000 TeV.

\section{\label{sec:conclusion}Conclusion}

In this work, we demonstrate that the latent space of the AE can be used as an input to reconstruct the parameters of the primary particle of the EAS. Using the example of determining the energy of the primary particle for TAIGA-HiSCORE data, we determined that the latent space dimensionality of 12 provides an optimal balance between model complexity and reconstruction accuracy. The standard deviation of the relative energy reconstruction error is 19--24\% for energies from 100 to 400 TeV and 13--15\% for energies from 400 to 1000 TeV, which is comparable with the results obtained by the conventional technique. This is a preliminary but promising result; we expect that increasing the training set and refining the hyperparameters will allow us to further improve the accuracy of energy reconstruction within the proposed model. We also expect that the developed technique will be useful in joint analysis of multimodal data from TAIGA-HiSCORE and TAIGA-IACTs. In our future work, we will focus on expanding the set of EAS parameters determined from the AE latent space using subsequent neural networks and on validating the method on other types of primary particles, particularly protons.

\section*{Acknowledgement}
The work was carried out using equipment provided by Moscow University within the framework of the Moscow State University Development Program and data obtained at the unique TAIGA experimental facility.

\section*{Funding}
This work was funded by the Russian Science Foundation (RSF), grant No. 24-11-00136, \url{https://rscf.ru/project/24-11-00136/}.

\end{document}